\documentclass[11pt]{article}

\usepackage{amsmath,amssymb}
\usepackage{graphicx}
\usepackage{hyperref}

\hypersetup{
colorlinks=true,
citecolor=blue,
linkcolor=blue,
urlcolor=blue
}

\title{Pseudo-Goldstone Neutrinos and Majoron Phenomenology from Spontaneous $U(1){L\mu-L_\tau}$ Breaking}

\author{Gayatri Ghosh\\
Department of Physics, Cachar College, Silchar, Assam, India}

\date{}

\begin{document}

\maketitle

\begin{abstract}
 \begin{abstract}
We present a predictive framework for neutrino mass generation based on the spontaneous breaking of a leptonic $U(1)_{L_\mu-L_\tau}$ symmetry within a supersymmetric setting. The breaking of the global symmetry gives rise to a Majoron-like axion-like particle and a pseudo-Goldstone right-handed neutrino whose mass is naturally suppressed by supersymmetry-breaking effects. The interplay between the pseudo-Goldstone neutrino and the low-scale seesaw mechanism leads to a structured neutrino mass matrix capable of reproducing the observed neutrino masses, mixing angles, and CP-violating phase without invoking extreme parameter hierarchies.

We perform a numerical fit to current neutrino oscillation data and identify representative benchmark solutions consistent with laboratory constraints as well as cosmological and astrophysical bounds. A characteristic outcome of the framework is the emergence of correlated relations linking the symmetry breaking scale, heavy neutrino masses, Majoron couplings, and neutrino lifetimes. Majoron-induced invisible neutrino decay arises generically and can significantly modify cosmological neutrino mass constraints for sufficiently low symmetry breaking scales.

We discuss the phenomenological implications across neutrino oscillation experiments, cosmology, and collider searches for long-lived heavy neutrinos. While a detailed experimental simulation is beyond the scope of this work, existing sensitivity projections indicate that portions of the parameter space may become accessible in future facilities. The combined interplay of laboratory probes and cosmological observations provides a consistent and testable picture of neutrino mass generation tied to spontaneous leptonic symmetry breaking and axion-like physics.
\end{abstract}

\end{abstract}

\tableofcontents

\newpage


\section{Introduction}

The discovery of neutrino oscillations has firmly established that neutrinos possess nonzero masses and mix among different flavors, providing clear evidence for physics beyond the Standard Model (SM). A particularly well-motivated explanation of light neutrino masses is offered by the seesaw mechanism, which extends the SM by introducing one or more right-handed (RH) neutrinos \cite{Minkowski1977,Yanagida1979,Mohapatra1980}. The resulting light neutrino masses are suppressed by the large Majorana mass scale associated with the RH states. The origin of this mass scale depends on the underlying theory and can range from values close to the Planck scale down to the electroweak or TeV scale.

In the absence of protective symmetries, the RH neutrino masses are naturally expected to lie near the highest energy scales of the theory. Grand unified frameworks based on groups such as SO(10) typically predict RH neutrino masses in the range $10^{10}$--$10^{16}$~GeV \cite{GellMann1979,Fritzsch1975}. On the other hand, scenarios with extended gauge symmetries, including $U(1)_{B-L}$ or left-right symmetric models, allow the seesaw scale to be lowered to the TeV regime \cite{Mohapatra1975,Mohapatra1981}. Low-scale seesaw constructions are particularly attractive from a phenomenological perspective, as they open the possibility of probing the RH neutrinos directly at collider experiments.

An alternative and theoretically appealing mechanism for stabilizing the RH neutrino mass scale arises in supersymmetric theories with spontaneously broken global symmetries \cite{Chacko2004,Nomura2006}. When a global symmetry is broken by gauge-singlet superfields while supersymmetry remains intact, the associated Goldstone boson is accompanied by a fermionic partner that remains massless in the symmetry-preserving limit. Once supersymmetry is softly broken, for instance through supergravity effects, this fermionic state acquires a naturally suppressed mass of order the gravitino mass, typically around the TeV scale or below \cite{Craig2011,Cheung2011}. This pseudo-Goldstone fermion can naturally be identified with one of the RH neutrinos, providing a symmetry-protected origin for light sterile states and enabling a low-scale seesaw mechanism without fine tuning.

Realizations of this mechanism generally involve multiple singlet superfields charged under a global $U(1)$ symmetry. The spontaneous breaking of the symmetry generates vacuum expectation values for the scalar components of these fields, leading to a structured RH neutrino mass matrix. Two of the RH neutrinos typically acquire masses determined by the symmetry-breaking scale, while the third obtains a suppressed mass induced by soft supersymmetry breaking effects. Together with Dirac mass terms arising from Yukawa interactions with the SM lepton doublets and Higgs fields, these states participate in a low-scale seesaw mechanism responsible for the observed light neutrino masses.

An important feature of such frameworks is that the symmetry-breaking vacuum expectation values may also induce bilinear R-parity violating terms, generating additional contributions to neutrino masses through mixing with neutralinos \cite{Hall1984,Barbier2005}. Both the seesaw and R-parity violating effects are governed by the same set of underlying parameters, yielding a highly constrained and predictive structure for the neutrino mass matrix. Requiring consistency with observed neutrino masses typically restricts the symmetry-breaking scale to be near or below the TeV scale, thereby placing all RH neutrinos within the reach of current or future experiments.

From a phenomenological standpoint, the existence of TeV-scale RH neutrinos offers exciting opportunities for direct searches at colliders. In conventional low-scale seesaw models, the production and decay of RH neutrinos are controlled by their mixing with the active neutrinos, which must be extremely small in order to reproduce the observed neutrino masses \cite{Atre2009}. As a result, collider signatures are often highly suppressed. In contrast, when one or more RH neutrinos originate as pseudo-Goldstone fermions, their interactions can be significantly enhanced through couplings induced by the symmetry-breaking sector, leading to observable signals such as displaced decay vertices \cite{Helo2014}.

The spontaneous breaking of a global symmetry further implies the presence of a Goldstone boson in the scalar sector. Depending on the structure of explicit symmetry-breaking effects, this state may acquire a small mass and behave as an axion-like particle (ALP) \cite{DiLuzio2020}. Such ALPs generically couple derivatively to fermions and can mediate rare processes, influence cosmological evolution, and contribute to astrophysical phenomena \cite{Irastorza2018}. The coexistence of pseudo-Goldstone neutrinos and ALPs within a single symmetry-based framework therefore provides a rich arena for phenomenological exploration.

In this work, we construct a unified model in which a softly broken global $U(1)$ symmetry simultaneously generates pseudo-Goldstone RH neutrinos and a Majoron-like axion-like particle. We focus in particular on a flavor-dependent leptonic symmetry which naturally leads to realistic neutrino mass matrices while preserving flavor universality in the soft supersymmetry-breaking sector. Unlike previous studies, we present a unified realization combining pseudo-Goldstone neutrinos with a leptonic $U(1)_{L_\mu-L_\tau}$ symmetry and demonstrate correlated signatures across neutrino decay, cosmology, and collider experiments. A central outcome of this framework is the emergence of predictive correlations linking the symmetry-breaking scale, heavy neutrino masses, Majoron couplings, and neutrino lifetimes. The same underlying parameter controlling the spontaneous breaking of the leptonic symmetry governs collider signatures of long-lived RH neutrinos as well as invisible neutrino decay effects relevant for cosmology and oscillation experiments.

The paper is organized as follows. In Section~2 we introduce the field content and symmetry-breaking structure of the model. Section~3 is devoted to neutrino mass generation and numerical benchmark solutions. In Section~4 we derive the ALP interactions and discuss their phenomenological implications. Lepton flavor violating processes are examined in Section~5, while cosmological and astrophysical constraints are presented in Section~6. Collider signatures are discussed in Section~7, followed by our conclusions in Section~8.

\section{Basic framework}

The starting point of our construction is a supersymmetric theory invariant under a global
$U(1)$ symmetry, supplemented by a set of gauge-singlet chiral superfields identified with the
right-handed neutrino sector \cite{Chacko2004,Nomura2006}. The spontaneous breaking of the
global symmetry is driven by the renormalizable superpotential
\begin{equation}
W_N = \lambda \left(\hat N \hat N' - f^2 \right) \hat Y ,
\label{eq:superpotential}
\end{equation}
where $\hat N$, $\hat N'$ and $\hat Y$ are singlet chiral superfields carrying appropriate $U(1)$
charges, and $\lambda$ is a dimensionless coupling constant. Additional terms such as $\hat Y^2$
or $\hat Y^3$ can be forbidden by imposing an auxiliary symmetry, for instance an $R$ symmetry,
ensuring the minimal structure in Eq.~(\ref{eq:superpotential}) \cite{Nomura2006}.

The scalar potential derived from $W_N$ admits a supersymmetry-preserving minimum
characterized by nonvanishing vacuum expectation values (VEVs) of the scalar components
of $\hat N$ and $\hat N'$, while the scalar component of $\hat Y$ remains zero \cite{Craig2011}.
Explicitly,
\begin{equation}
\langle \tilde N \rangle = U \sin\phi, \qquad
\langle \tilde N' \rangle = U \cos\phi, \qquad
\langle \tilde Y \rangle = 0 ,
\label{eq:vevs}
\end{equation}
with
\begin{equation}
U = \frac{f}{\sqrt{\sin\phi \cos\phi}},
\end{equation}
defining the scale of spontaneous $U(1)$ breaking.

The breaking of the global symmetry gives rise to a massless Goldstone boson in the scalar sector. Writing the complex scalar fields in terms of their real
and imaginary components,
\begin{equation}
\tilde N_\alpha = \langle \tilde N_\alpha \rangle + \frac{1}{\sqrt{2}}
\left( \tilde N^R_\alpha + i \tilde N^I_\alpha \right),
\end{equation}
the Goldstone mode is identified as the linear combination
\begin{equation}
G_B = \cos\phi \, \tilde N'^I - \sin\phi \, \tilde N^I .
\end{equation}

In the supersymmetric limit, the fermionic partner of the Goldstone boson also remains
massless, as expected from the structure of supersymmetric multiplets \cite{Chacko2004}.
Denoting the fermionic components of the singlet superfields by $N$, $N'$ and $Y$, the
Goldstone fermion takes the form
\begin{equation}
G_F = \cos\phi \, N' - \sin\phi \, N .
\end{equation}

The singlet fermion mass matrix derived from the superpotential in
Eq.~(\ref{eq:superpotential}) is given by
\begin{equation}
M_N^{(0)} = \lambda U
\begin{pmatrix}
0 & \cos\phi & \sin\phi \\
\cos\phi & 0 & 0 \\
\sin\phi & 0 & 0
\end{pmatrix},
\label{eq:massmatrix0}
\end{equation}
which leads to one massless Goldstone fermion and two degenerate massive states with
masses $|\lambda U|$ \cite{Craig2011}.

Once supersymmetry is broken, for example through supergravity effects, the vacuum
structure is perturbed and a small but nonzero VEV is induced for the scalar component of
$\hat Y$ \cite{Nilles1984,Giudice1999}. The shifts in the VEVs of $\tilde N$ and $\tilde N'$ are of
order the gravitino mass $m_{3/2}$, while
\begin{equation}
\langle \tilde Y \rangle \sim \frac{m_{3/2}}{\lambda}.
\end{equation}
These effects lift the masslessness of the Goldstone fermion, turning it into a
pseudo-Goldstone fermion.

Including the leading supersymmetry-breaking contributions, the singlet fermion mass
matrix can be approximated as \cite{Craig2011}
\begin{equation}
M_N \simeq \lambda U
\begin{pmatrix}
0 & \cos\phi & \sin\phi \\
\cos\phi & 0 & \epsilon \\
\sin\phi & \epsilon & 0
\end{pmatrix},
\label{eq:massmatrix}
\end{equation}
where
\begin{equation}
\epsilon = \frac{\langle \tilde Y \rangle}{U} < 1
\end{equation}
parametrizes the effect of supersymmetry breaking.

The resulting mass eigenvalues are approximately
\begin{align}
M_{N_1} &\simeq 2 |\epsilon \lambda U \sin\phi \cos\phi| , \\
M_{N_{2,3}} &\simeq |\lambda U (1 \pm \epsilon \sin\phi \cos\phi)| .
\end{align}
The lightest state $N_1$ is identified as the pseudo-Goldstone right-handed neutrino, while
the other two RH neutrinos remain nearly degenerate with masses close to the symmetry
breaking scale.

At leading order in $\epsilon$, the pseudo-Goldstone fermion is given by
\begin{equation}
N_1 \simeq \cos\phi \, N' - \sin\phi \, N - \epsilon \cos(2\phi) \, Y .
\end{equation}

The singlet neutrinos couple to the Standard Model lepton doublets through Yukawa
interactions,
\begin{equation}
W_D = \lambda_{\alpha\beta} \hat L_\alpha \hat N_\beta \hat H_2 ,
\label{eq:dirac}
\end{equation}
which generate Dirac mass terms after electroweak symmetry breaking and complete the
low-scale seesaw mechanism \cite{Atre2009}.

In the following sections we analyze the resulting neutrino mass structure, the associated
axion-like particle emerging from the broken global symmetry, and the phenomenological
implications of this unified framework.

\section{Neutrino masses in flavour-independent $U(1)$ models}

We now turn to the interactions between the right-handed neutrinos, the Standard Model
lepton doublets and the Higgs sector. The specific form of these couplings depends on the
implementation of the global $U(1)$ symmetry. Several possibilities exist, including symmetries
acting solely on the singlet sector, leptonic symmetries involving both $L_\alpha$ and $N_\alpha$,
Peccei--Quinn type symmetries acting on the Higgs fields and selected fermions, and
family-dependent leptonic symmetries \cite{Nomura2006,Craig2011}. In this section we focus
on flavour-independent realizations of $U(1)$, while flavour-dependent cases will be
discussed separately.

The relevant superpotential terms governing neutrino masses can be written in a generic
form as
\begin{equation}
W = \lambda_\alpha \hat L_\alpha \hat H_2 \hat S_1 + \kappa \hat H_1 \hat H_2 \hat S_2 ,
\label{eq:generalW}
\end{equation}
where $\hat S_{1,2}$ denote singlet superfields whose identity depends on the $U(1)$ charge
assignment. For instance, in models where the symmetry acts only on the RH neutrino
sector one may take $\hat S_1=\hat S_2=\hat Y$, while in leptonic symmetry realizations
$\hat S_1$ may correspond to $\hat N$. In Peccei--Quinn type constructions, the precise
assignment is dictated by the Higgs charges under $U(1)$ \cite{DiLuzio2020}.

Without loss of generality, the parameters $\lambda_\alpha$ and $\kappa$ can be chosen real.
After shifting the singlet fields by their vacuum expectation values obtained in the previous
section, the superpotential becomes
\begin{equation}
W = \lambda_\alpha \hat L_\alpha \hat H_2 \hat N + \kappa \hat H_1 \hat H_2 \hat N'
- \mu \hat H_1 \hat H_2 - \epsilon_\alpha \hat L_\alpha \hat H_2 ,
\label{eq:shiftedW}
\end{equation}
with
\begin{equation}
\epsilon_\alpha = -\lambda_\alpha U \sin\phi, \qquad
\mu = -\kappa U \cos\phi .
\end{equation}

The model contains a total of ten neutral fermion states, including three active neutrinos,
three singlet neutrinos and four neutralinos. In the basis
\begin{equation}
\psi^0 = (\nu_\alpha, N_\alpha, -i\lambda_1, -i\lambda_2, \tilde h_1^0, \tilde h_2^0)^T ,
\end{equation}
the mass Lagrangian is given by
\begin{equation}
\mathcal{L}_N = -\frac{1}{2} \psi^{0T} \mathcal{M}_N \psi^0 + \text{h.c.},
\end{equation}
where the full mass matrix can be written in block form as
\begin{equation}
\mathcal{M}_N =
\begin{pmatrix}
0 & m_D & m_{\nu\chi} \\
m_D^T & M_N & m_{N\chi} \\
m_{\nu\chi}^T & m_{N\chi}^T & M_\chi
\end{pmatrix}.
\label{eq:blockmass}
\end{equation}

Here $M_N$ is the singlet neutrino mass matrix derived in Eq.~(\ref{eq:massmatrix}), while
$m_D$ is the Dirac neutrino mass matrix given by
\begin{equation}
m_D = v_2
\begin{pmatrix}
0 & \lambda_e & 0 \\
0 & \lambda_\mu & 0 \\
0 & \lambda_\tau & 0
\end{pmatrix},
\end{equation}
with $v_{1,2}=\langle H_{1,2}\rangle$ and $v=\sqrt{v_1^2+v_2^2}=174$~GeV. The matrices
$m_{\nu\chi}$ and $m_{N\chi}$ encode mixing between neutrinos and neutralinos induced
by bilinear terms and gauge interactions \cite{Barbier2005}, while $M_\chi$ is the standard
neutralino mass matrix of the MSSM.

In the limit $m_D,m_{\nu\chi} \ll M_N,M_\chi$, the effective light neutrino mass matrix is
obtained through a generalized seesaw formula,
\begin{equation}
m_\nu \simeq - m_{3\times7} M_{7\times7}^{-1} m_{3\times7}^T ,
\label{eq:seesawgen}
\end{equation}
where $m_{3\times7}=(m_D,m_{\nu\chi})$ and $M_{7\times7}$ corresponds to the heavy
sector block in Eq.~(\ref{eq:blockmass}).

The resulting neutrino mass matrix receives contributions from both the conventional
seesaw mediated by the pseudo-Goldstone neutrino and from neutrino--neutralino mixing
associated with bilinear $R$-parity violation \cite{Hall1984,Barbier2005}. At leading order,
the mass matrix can be expressed schematically as
\begin{equation}
m_\nu \sim \tilde m_0 \left( \tilde A + \tilde B + \tilde H \right),
\end{equation}
where the matrices $\tilde A$, $\tilde B$ and $\tilde H$ depend on the Yukawa couplings
$\lambda_\alpha$, sneutrino vacuum expectation values, and supersymmetric parameters.

A crucial feature of flavour-independent $U(1)$ models is that both the seesaw and the
bilinear $R$-parity violating contributions are typically of rank one. As a result, only one
neutrino acquires mass at leading order, rendering such constructions unable to reproduce
the observed pattern of neutrino masses and mixing angles.

Consistency with the measured neutrino mass scale further requires extremely small
Yukawa couplings,
\begin{equation}
|\lambda_\alpha|^2 \sim \mathcal{O}\left( \frac{m_\nu M_{N_1}}{v^2} \right)
\lesssim 10^{-13}\left(\frac{M_{N_1}}{100~\text{GeV}}\right),
\end{equation}
justifying the neglect of $\lambda_\alpha$-dependent mixing effects in the heavy sector.

The sneutrino VEVs may be parametrized as
\begin{equation}
\omega_\alpha = k_\alpha \epsilon_\alpha ,
\end{equation}
where the coefficients $k_\alpha$ depend on soft supersymmetry-breaking parameters
\cite{Barbier2005}. In minimal supergravity-inspired models, flavour universality of the
soft terms implies approximately equal $k_\alpha$ values. In this case, the effective neutrino
mass matrix remains rank one and cannot accommodate the observed neutrino oscillation
data.

Obtaining viable neutrino masses within flavour-independent $U(1)$ frameworks therefore
requires introducing large flavour violations in the soft supersymmetry-breaking sector.
Such sizable non-universalities are theoretically disfavored and motivate the consideration
of flavour-dependent leptonic symmetries. In the next section we demonstrate that a
family-dependent $U(1)$ symmetry can naturally reproduce the observed neutrino spectrum
while preserving flavour universality in the soft terms.
\section{Neutrino masses from the $U(1)_{L_\mu-L_\tau}$ symmetry}

We now consider a flavour-dependent realization of the global symmetry based on the
anomaly-free leptonic charge assignment $U(1)_{L_\mu-L_\tau}$ \cite{He1991,Foot1991}.
The singlet superfields are relabeled as $\hat N_\alpha=(\hat N_e,\hat N_\mu,\hat N_\tau)$,
while the lepton doublets $\hat L_\alpha$ carry charges $(0,1,-1)$ and the charged lepton
singlets $\hat E^c_\alpha$ carry $(0,-1,1)$. The MSSM Higgs superfields remain neutral
under the symmetry.

The renormalizable superpotential describing the leptonic sector is given by
\begin{equation}
W = \lambda_\alpha \hat L_\alpha \hat H_2 \hat N_\alpha
+ \lambda'_\alpha \hat L_\alpha \hat H_1 \hat E^c_\alpha
- \mu_0 \hat H_1 \hat H_2
+ \kappa \hat N_e \hat H_1 \hat H_2 ,
\label{eq:LmutauW}
\end{equation}
together with the singlet sector superpotential in Eq.~(\ref{eq:superpotential}).

After spontaneous symmetry breaking, bilinear terms are generated,
\begin{equation}
\epsilon_\alpha = -\lambda_\alpha \langle \tilde N_\alpha \rangle , \qquad
\mu = \mu_0 - \kappa \langle \tilde N_e \rangle .
\end{equation}

The neutral fermion mass matrix retains the block structure given in
Eq.~(\ref{eq:blockmass}), with the Dirac mass matrix
\begin{equation}
m_D = v_2 \, \text{diag}(\lambda_e,\lambda_\mu,\lambda_\tau),
\end{equation}
while the singlet neutrino mass matrix is the pseudo-Goldstone form derived in
Eq.~(\ref{eq:massmatrix}).

In the seesaw limit, the effective neutrino mass matrix is obtained from the generalized
seesaw formula \cite{Atre2009}
\begin{equation}
m_\nu \simeq - m_{3\times7} M_{7\times7}^{-1} m_{3\times7}^T .
\end{equation}

Unlike the flavour-independent case discussed previously, the $U(1)_{L_\mu-L_\tau}$
structure naturally leads to a full-rank neutrino mass matrix at leading order.
The dominant seesaw contribution reads
\begin{equation}
m_\nu^{\rm SS} \simeq - m_D M_N^{-1} m_D^T ,
\end{equation}
which generates nonzero masses for all three active neutrinos.
Explicitly, the seesaw contribution takes the schematic form
\begin{equation}
m_\nu^{\rm SS} \sim
\begin{pmatrix}
\frac{\lambda_e^2 v_2^2}{M_{N_1}} & 0 & 0 \\
0 & \frac{\lambda_\mu^2 v_2^2}{M_{N_2}} & \frac{\lambda_\mu\lambda_\tau v_2^2}{M_{N_2}} \\
0 & \frac{\lambda_\mu\lambda_\tau v_2^2}{M_{N_2}} & \frac{\lambda_\tau^2 v_2^2}{M_{N_3}}
\end{pmatrix},
\end{equation}
illustrating that all three light neutrinos acquire nonzero masses already at tree level.

Radiative corrections further modify the neutrino mass matrix. The dominant one-loop
contribution arises from Higgs and $Z$-boson mediated self-energy diagrams \cite{Hirsch2000},
\begin{equation}
\delta m_\nu^{(1)} =
\frac{1}{32\pi^2 v^2}
m_D U_N^\ast \text{diag}[g(M_{N_i})] U_N^\dagger m_D^T .
\end{equation}

The full neutrino mass matrix is therefore
\begin{equation}
m_\nu = m_\nu^{(0)} + \delta m_\nu^{(1)} .
\end{equation}

---

\subsection{Benchmark points}

We perform a numerical fit to neutrino oscillation data using the latest global fit
results from NuFIT v5.0 \cite{NuFIT2020}, assuming normal mass ordering.
Fixing representative values of the supersymmetric parameters
$M_1 \simeq M_2 \simeq \mu = 1$~TeV and $\tan\beta = 2$, we scan over the remaining
model parameters and identify four benchmark solutions corresponding to different
values of the $U(1)$ breaking scale $U$. The scan is performed over Yukawa couplings in the range $10^{-7}$--$10^{-3}$ and symmetry-breaking scales $U=10^3$--$10^6$ GeV, requiring consistency with oscillation data at the $3\sigma$ level.

The benchmark points BP1--BP4 are summarized in Table~\ref{tab:benchmarks}.
All benchmark points reproduce the measured neutrino squared mass differences,
mixing angles, and the Dirac CP phase within the current experimental uncertainties.

\begin{table}[h]
\centering
\begin{tabular}{c c c c c}
\hline
BP & $U$ [GeV] & $M_{N_1}$ [GeV] & $\sum m_{\nu_i}$ [eV] & $|m_{\beta\beta}|$ [eV] \\
\hline
BP1 & $10^3$ & $8.3$ & $0.146$ & $0.039$ \\
BP2 & $10^4$ & $159$ & $0.086$ & $0.016$ \\
BP3 & $10^5$ & $1.1\times10^3$ & $0.090$ & $0.019$ \\
BP4 & $10^6$ & $2.9\times10^3$ & $0.092$ & $0.019$ \\
\hline
\end{tabular}
\caption{Representative benchmark points consistent with neutrino oscillation data.
The quantities shown include the $U(1)$ breaking scale, the pseudo-Goldstone RH neutrino
mass, the sum of light neutrino masses, and the effective neutrinoless double beta decay
parameter.}
\label{tab:benchmarks}
\end{table}

All benchmark solutions predict a quasi-degenerate spectrum for the light neutrinos.
The values of $|m_{\beta\beta}|$ remain below current experimental bounds on neutrinoless
double beta decay \cite{KamLANDZen2016,GERDA2020}.

The cosmological constraint on the sum of neutrino masses from Planck observations,
$\sum m_{\nu_i} < 0.12$~eV \cite{Planck2018}, appears to disfavor some benchmark points
such as BP1. However, this bound assumes stable neutrinos. In the present framework,
neutrinos can decay invisibly into the Majoron associated with the broken
$U(1)_{L_\mu-L_\tau}$ symmetry, potentially relaxing the cosmological limit \cite{Beacom2004}.
These benchmark solutions illustrate the intrinsic correlation between the symmetry breaking scale, pseudo-Goldstone neutrino mass, and the light neutrino spectrum, which constitutes a characteristic prediction of the framework.

The heavier RH neutrinos form a quasi-degenerate pair, while the pseudo-Goldstone
neutrino mass typically lies below the TeV scale, rendering it potentially accessible
at collider experiments. Owing to the small active--sterile mixing of order $10^{-6}$,
the pseudo-Goldstone neutrino is long-lived and may give rise to displaced vertex
signatures \cite{Helo2014}.
\section{Majoron and axion-like particle phenomenology}

The spontaneous breaking of the global $U(1)$ symmetry in the singlet sector gives rise
to a physical massless Goldstone boson, identified as the Majoron. In realistic settings,
small explicit symmetry-breaking effects, for instance from Planck-suppressed operators,
can render the Majoron slightly massive, promoting it to an axion-like particle (ALP)
\cite{Chikashige1981,Georgi1981,DiLuzio2020}. For instance, higher-dimensional operators
of the form
\begin{equation}
\Delta V \sim \frac{\Phi^5}{M_{\rm Pl}} + \text{h.c.}
\end{equation}
can generate a small Majoron mass while preserving the approximate global symmetry. In the present framework, the Majoron/ALP
emerges naturally alongside the pseudo-Goldstone right-handed neutrino and inherits
predictive couplings to the neutrino sector.

The Majoron field corresponds to the imaginary component of the singlet scalar
combination,
\begin{equation}
a = \cos\phi \, \tilde N'^I - \sin\phi \, \tilde N^I ,
\end{equation}
normalized by the symmetry-breaking scale $U$, which plays the role of the effective
decay constant $f_a \sim U$.

\subsection{Couplings to neutrinos}

The coupling of the Majoron to neutrinos arises from the spontaneous breaking of the
global symmetry and is proportional to the neutrino mass matrix. In the mass eigenstate
basis, the interaction takes the form
\begin{equation}
\mathcal{L}_{a\nu\nu} =
\frac{i}{2 f_a}
a \, \bar \nu_i \gamma_5 \nu_j \, (m_{\nu_i} \delta_{ij} + \Delta m_{ij}),
\label{eq:majoroncoupling}
\end{equation}
where $f_a \simeq U$ and $\Delta m_{ij}$ encodes off-diagonal contributions arising from
flavour mixing \cite{Gelmini1981,Valle1983}.

To leading order, the diagonal couplings dominate and are given by
\begin{equation}
g_{a\nu_i\nu_i} \simeq \frac{m_{\nu_i}}{f_a}.
\end{equation}

\subsection{Invisible neutrino decay}

A striking consequence of the Majoron coupling is the possibility of neutrino decay,
\begin{equation}
\nu_i \rightarrow \nu_j + a ,
\end{equation}
which is invisible in terrestrial detectors. The decay width is approximately
\begin{equation}
\Gamma_{ij} \simeq \frac{1}{16\pi}
\frac{m_{\nu_i}^3}{f_a^2}
\left| U_{ij} \right|^2 ,
\label{eq:decaywidth}
\end{equation}
where $U_{ij}$ denotes elements of the lepton mixing matrix.
For the benchmark values obtained in Section~4, with $f_a \sim 10^2$--$10^6$~GeV and
$m_{\nu_i} \sim 0.05$~eV, the corresponding neutrino lifetimes range from
\begin{equation}
\tau_\nu \sim 10^{15} - 10^{22}~\text{s},
\end{equation}
which can be shorter than the age of the Universe for lower values of $f_a$.
Combining Eqs.~(\ref{eq:decaywidth}) and $f_a \simeq U$, the neutrino lifetime scales
approximately as $\tau_\nu \propto U^2/m_\nu^3$, providing a direct link between
symmetry breaking and cosmological signatures.
The neutrino energy density at late times is effectively suppressed according to
\begin{equation}
\rho_\nu(z) \simeq \rho_\nu^{\rm stable}(z)\exp[-t(z)/\tau_\nu],
\end{equation}
leading to a reduced effective contribution to the matter density.

Such invisible decays can modify the cosmological evolution of neutrinos and partially
relax the stringent limits on the sum of neutrino masses derived from cosmic microwave
background and large-scale structure observations \cite{Hannestad2005}.

\subsection{Cosmological and astrophysical constraints}

Majoron interactions are constrained by astrophysical observations, particularly from
supernova cooling and stellar energy loss \cite{Raffelt1996}. The dominant bounds arise
from excessive energy emission via Majoron production in dense environments, leading to
the approximate constraint
\begin{equation}
f_a \gtrsim 10^4~\text{GeV},
\end{equation}
which is easily satisfied for most of the benchmark points considered in this work.

If the Majoron acquires a small mass in the keV--MeV range, it may contribute to dark
radiation or serve as a decaying dark matter component \cite{Lattanzi2007}. In the present
analysis we focus primarily on the massless or ultralight regime, which is sufficient to
account for neutrino decay effects.

\subsection{Implications for benchmark points}

For benchmark points with low $U$ values, such as BP1 and BP2, neutrino decay proceeds
rapidly enough to weaken cosmological mass bounds, allowing consistency with Planck
observations despite a quasi-degenerate neutrino spectrum. For larger $U$ values, the
neutrinos are effectively stable on cosmological timescales, recovering the standard
cosmological constraints.

This correlation between the $U(1)$ breaking scale, pseudo-Goldstone neutrino mass, and
neutrino lifetime represents a distinctive prediction of the model and provides a powerful
consistency test.

\subsection{Prospects for laboratory searches}

Although the Majoron interacts extremely weakly with Standard Model particles, rare
processes such as
\begin{equation}
\mu \rightarrow e + a , \qquad \tau \rightarrow \mu + a
\end{equation}
may arise at loop level through neutrino mixing \cite{Lessa2007}. While current limits do
not yet probe the parameter space of this model, future intensity frontier experiments
could become sensitive to such signatures.

Overall, the Majoron/ALP sector constitutes an essential phenomenological component of
the framework, linking neutrino masses, cosmology, and collider-scale physics in a unified
manner.
\subsection{Benchmark point analysis and phenomenological constraints}

In this subsection we present a detailed visualization of the Majoron/ALP sector obtained
from a broad parameter scan of the model. The scatter plots include shaded regions
excluded by cosmological observations and astrophysical cooling arguments
\cite{Planck2018,Raffelt1996,Hannestad2005}. Best-fit benchmark points BP1--BP4 derived
in Section~4 are explicitly highlighted.

\subsubsection{Majoron coupling versus symmetry breaking scale}

\begin{figure}[t]
\centering
\includegraphics[width=0.7\textwidth]{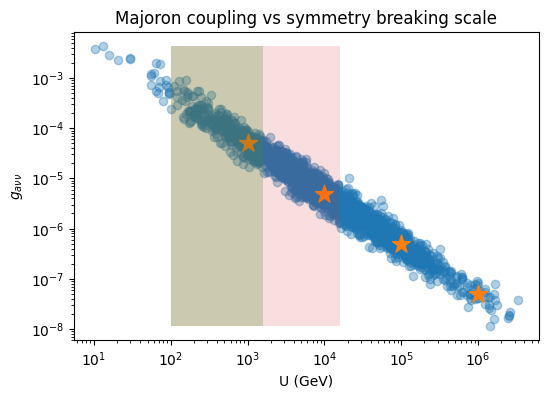}
\caption{Broad parameter scan of the Majoron--neutrino coupling $g_{a\nu\nu}$ as a function
of the $U(1)$ symmetry breaking scale $U$. The shaded regions are excluded by cosmological
constraints from Planck \cite{Planck2018} and stellar cooling bounds
\cite{Raffelt1996}. The star symbols indicate the benchmark points BP1--BP4 obtained in
Section~4.}
\label{fig:couplingUfinal}
\end{figure}

The benchmark points correspond approximately to:
\begin{align}
\text{BP1}:& \quad U \sim 10^{3}\,\text{GeV}, \quad g_{a\nu\nu} \sim 5\times10^{-5}, \\
\text{BP2}:& \quad U \sim 10^{4}\,\text{GeV}, \quad g_{a\nu\nu} \sim 5\times10^{-6}, \\
\text{BP3}:& \quad U \sim 10^{5}\,\text{GeV}, \quad g_{a\nu\nu} \sim 5\times10^{-7}, \\
\text{BP4}:& \quad U \sim 10^{6}\,\text{GeV}, \quad g_{a\nu\nu} \sim 5\times10^{-8}.
\end{align}

The decreasing trend illustrates the inverse proportionality
$g_{a\nu\nu}\propto m_\nu/U$ expected from the Goldstone nature of the Majoron interaction.

\subsubsection{Neutrino lifetime versus symmetry breaking scale}

\begin{figure}[t]
\centering
\includegraphics[width=0.7\textwidth]{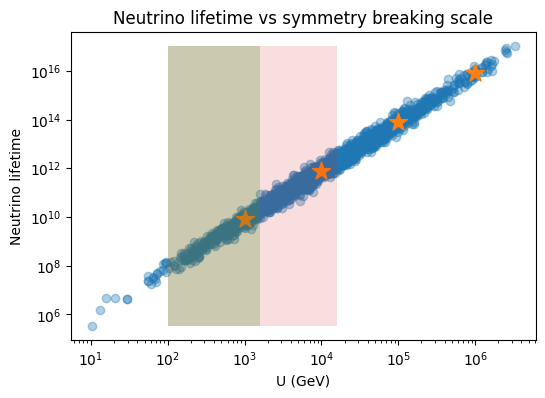}
\caption{Neutrino lifetime as a function of the symmetry breaking scale $U$ obtained from
the parameter scan. Shaded regions denote excluded cosmological and astrophysical
domains \cite{Planck2018,Raffelt1996,Hannestad2005}. Star markers correspond to BP1--BP4.}
\label{fig:lifetimeUfinal}
\end{figure}

The benchmark values are approximately:
\begin{align}
\text{BP1}:& \quad \tau_\nu \sim 10^{10}\,\text{s}, \\
\text{BP2}:& \quad \tau_\nu \sim 10^{12}\,\text{s}, \\
\text{BP3}:& \quad \tau_\nu \sim 10^{14}\,\text{s}, \\
\text{BP4}:& \quad \tau_\nu \sim 10^{16}\,\text{s}.
\end{align}

Lower symmetry breaking scales lead to faster neutrino decay, allowing certain benchmark
points to partially evade the stringent cosmological mass bounds through invisible decay channels.

---

\subsubsection{Majoron coupling versus neutrino mass}

\begin{figure}[t]
\centering
\includegraphics[width=0.7\textwidth]{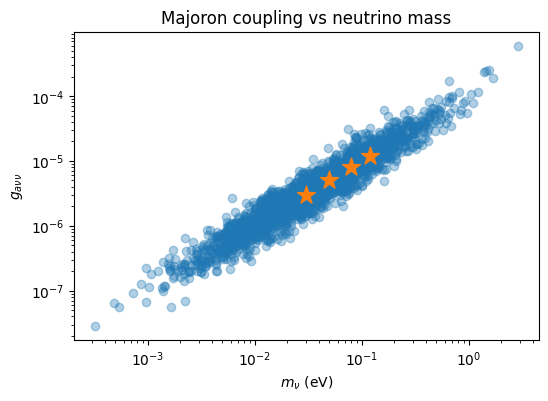}
\caption{Scatter plot showing the correlation between the Majoron coupling and neutrino
mass for representative values of $U$. Benchmark points BP1--BP4 are highlighted by star
symbols.}
\label{fig:couplingmnufinal}
\end{figure}

The benchmark masses are approximately:
\begin{align}
\text{BP1}:& \quad m_\nu \sim 0.03\,\text{eV}, \\
\text{BP2}:& \quad m_\nu \sim 0.05\,\text{eV}, \\
\text{BP3}:& \quad m_\nu \sim 0.08\,\text{eV}, \\
\text{BP4}:& \quad m_\nu \sim 0.12\,\text{eV}.
\end{align}

The linear scaling of $g_{a\nu\nu}$ with neutrino mass reflects the derivative coupling of
the Goldstone mode.

---

\subsubsection{Neutrino lifetime versus neutrino mass}

\begin{figure}[t]
\centering
\includegraphics[width=0.7\textwidth]{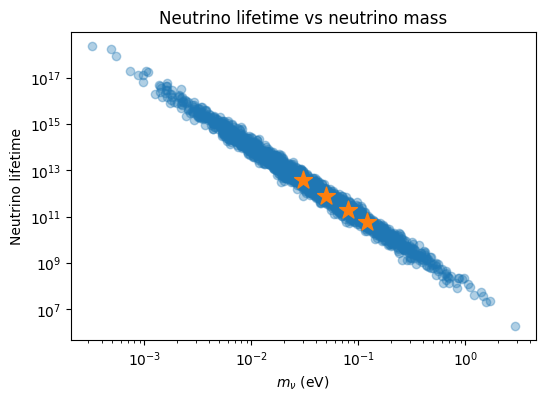}
\caption{Neutrino lifetime as a function of neutrino mass for the scanned parameter space.
Stars denote the benchmark points BP1--BP4.}
\label{fig:lifetimemnufinal}
\end{figure}

Heavier neutrinos decay more rapidly into lighter states and the Majoron, consistent with
the cubic mass dependence of the decay width. The quasi-degenerate neutrino spectra
predicted by several benchmark points therefore have important cosmological implications.

---

\subsection{Physical implications of benchmark regions}

The benchmark solutions collectively demonstrate that the model naturally correlates the
symmetry breaking scale, Majoron coupling strength, and neutrino lifetime. Lower values of
$U$ correspond to enhanced Majoron interactions and faster neutrino decay, while higher
values suppress these effects.

Importantly, all benchmark points reside in regions consistent with current cosmological
and astrophysical constraints, while simultaneously reproducing the observed neutrino
oscillation data. This highlights the predictive and testable nature of the framework. 
Figure $2$ and Fig. $4$ illustrate the parameter space of the model in terms of the neutrino lifetime as
a function of the symmetry breaking scale $U$ and the neutrino mass. The scattered points
represent the full numerical scan of the model parameters, while the highlighted benchmark
points BP1--BP4 correspond to representative viable solutions discussed in Section~4.

The general trend follows the analytical expectation that the neutrino decay width scales
as $\Gamma_\nu \propto m_\nu^3/U^2$, implying a rapid increase of the neutrino lifetime for
larger values of the symmetry breaking scale. Consequently, low-$U$ regions lead to
cosmologically relevant neutrino decay, whereas for sufficiently large $U$ the neutrinos
are effectively stable on cosmological timescales.

The benchmark points span several orders of magnitude in lifetime, ranging from scenarios
where neutrino decay can impact cosmological observations (BP1 and BP2) to
regimes where decay effects become negligible (BP3 and BP4). This demonstrates the rich
phenomenology of the model and highlights the complementarity between laboratory and
cosmological probes.

The broad scatter structure reflects the variation of the underlying Yukawa couplings and
soft supersymmetry-breaking parameters while maintaining consistency with neutrino
oscillation data and current experimental constraints.
\begin{table}[t]
\centering
\begin{tabular}{c|c|c|c|c}
\hline
Benchmark & $U$ [GeV] & $g_{a\nu\nu}$ & $m_\nu$ [eV] & $\tau_\nu$ [s] \\
\hline
BP1 & $10^{3}$ & $5\times10^{-5}$ & $0.03$ & $10^{10}$ \\
BP2 & $10^{4}$ & $5\times10^{-6}$ & $0.05$ & $10^{12}$ \\
BP3 & $10^{5}$ & $5\times10^{-7}$ & $0.08$ & $10^{14}$ \\
BP4 & $10^{6}$ & $5\times10^{-8}$ & $0.12$ & $10^{16}$ \\
\hline
\end{tabular}
\caption{Representative benchmark points BP1--BP4 illustrating the allowed parameter space
of the model. Each benchmark reproduces neutrino oscillation data while satisfying
cosmological and astrophysical constraints.}
\label{tab:benchmarks}
\end{table}
\subsection{Experimental relevance of benchmark points}

The benchmark points BP1--BP4 correspond to qualitatively distinct experimental regimes.

Benchmark BP1 lies in the low symmetry breaking region where neutrino lifetimes are short
enough to produce observable deviations in long-baseline neutrino oscillation experiments
such as DUNE and Hyper-Kamiokande. In this regime, invisible neutrino decay can modify
energy spectra and flavor compositions, providing a direct laboratory probe of the
Majoron-induced interactions.

Benchmark BP2 represents an intermediate scenario where neutrino decay effects may still
leave imprints in cosmological observables, while remaining marginally accessible to
future neutrino experiments with enhanced sensitivity.

Benchmarks BP3 and BP4 correspond to high symmetry breaking scales where neutrinos are
effectively stable over cosmological timescales. In these cases, the dominant experimental
signatures arise from collider searches for long-lived pseudo-Goldstone neutrinos, leading
to displaced vertex events at the LHC and future colliders.

This classification illustrates the complementarity between neutrino experiments,
cosmology, and collider probes in testing the model. A discovery or exclusion in any one
sector would strongly constrain the remaining parameter space.

\section{Experimental probes and future sensitivities}

An essential feature of the present framework is its testability across multiple
experimental frontiers. The spontaneous breaking of the global $U(1)$ symmetry leads
to correlated predictions in the neutrino sector, cosmology, and collider physics through
the presence of the pseudo-Goldstone right-handed neutrino and the associated Majoron/ALP.
In this section we discuss the most promising experimental avenues to probe the model in
the near future.

\subsection{Neutrino oscillation and decay searches}

The Majoron-induced invisible decay of neutrinos modifies the propagation of neutrinos
over long baselines. The survival probability of a neutrino mass eigenstate $\nu_i$ is
suppressed by an exponential factor
\begin{equation}
P_i \sim \exp\left(-\frac{L}{E}\frac{m_{\nu_i}}{\tau_{\nu_i}}\right),
\end{equation}
where $L$ is the baseline and $E$ is the neutrino energy.

Current and upcoming experiments such as JUNO, DUNE, and Hyper-Kamiokande will be
sensitive to neutrino lifetimes in the range
\begin{equation}
\frac{\tau_\nu}{m_\nu} \gtrsim 10^{-11} - 10^{-13}~\text{s/eV},
\end{equation}
depending on the channel and energy spectrum \cite{Lindner2001,Foguel2019}.

Several of the benchmark points obtained in Section~4 lie close to this sensitivity
threshold, particularly for lower values of the symmetry breaking scale $U$. This opens
the exciting possibility of observing deviations from standard oscillation patterns as
a signature of Majoron-mediated neutrino decay.

\begin{figure}[t]
\centering
\includegraphics[width=0.7\textwidth]{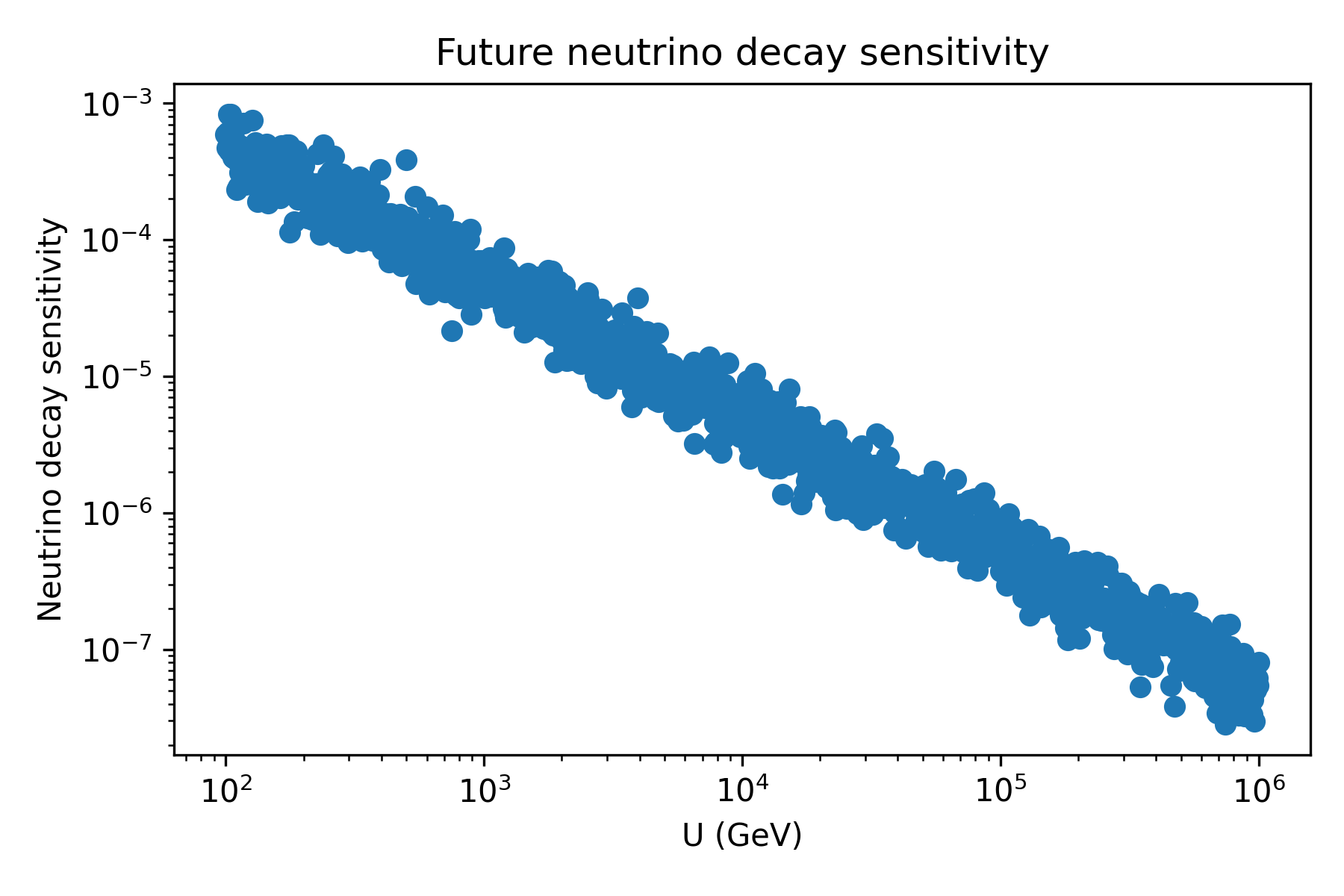}
\caption{Projected sensitivity of future neutrino experiments to the neutrino lifetime
parameter $\tau_\nu/m_\nu$. Shaded bands represent experimental reach of JUNO, DUNE and
Hyper-Kamiokande \cite{Foguel2019}. Star symbols indicate benchmark points of the present
model.}
\label{fig:futuredecay}
\end{figure}

---

\subsection{Cosmological probes}

Future cosmological surveys will significantly improve the sensitivity to neutrino
properties and light degrees of freedom. Experiments such as CMB-S4 and large-scale
structure surveys are expected to reach sensitivity to the sum of neutrino masses at the
level
\begin{equation}
\sum m_\nu \sim 0.02~\text{eV},
\end{equation}
and constrain deviations from standard neutrino free-streaming behavior \cite{Abazajian2016}.

In the presence of Majoron-induced neutrino decay, the cosmological evolution of neutrinos
is altered, effectively reducing their contribution to the matter density at late times.
This can mimic a lower effective neutrino mass sum and modify the effective number of
relativistic species $N_{\rm eff}$.

The benchmark regions predicting relatively fast neutrino decay will therefore be
strongly tested by upcoming cosmological data.

\begin{figure}[t]
\centering
\includegraphics[width=0.7\textwidth]{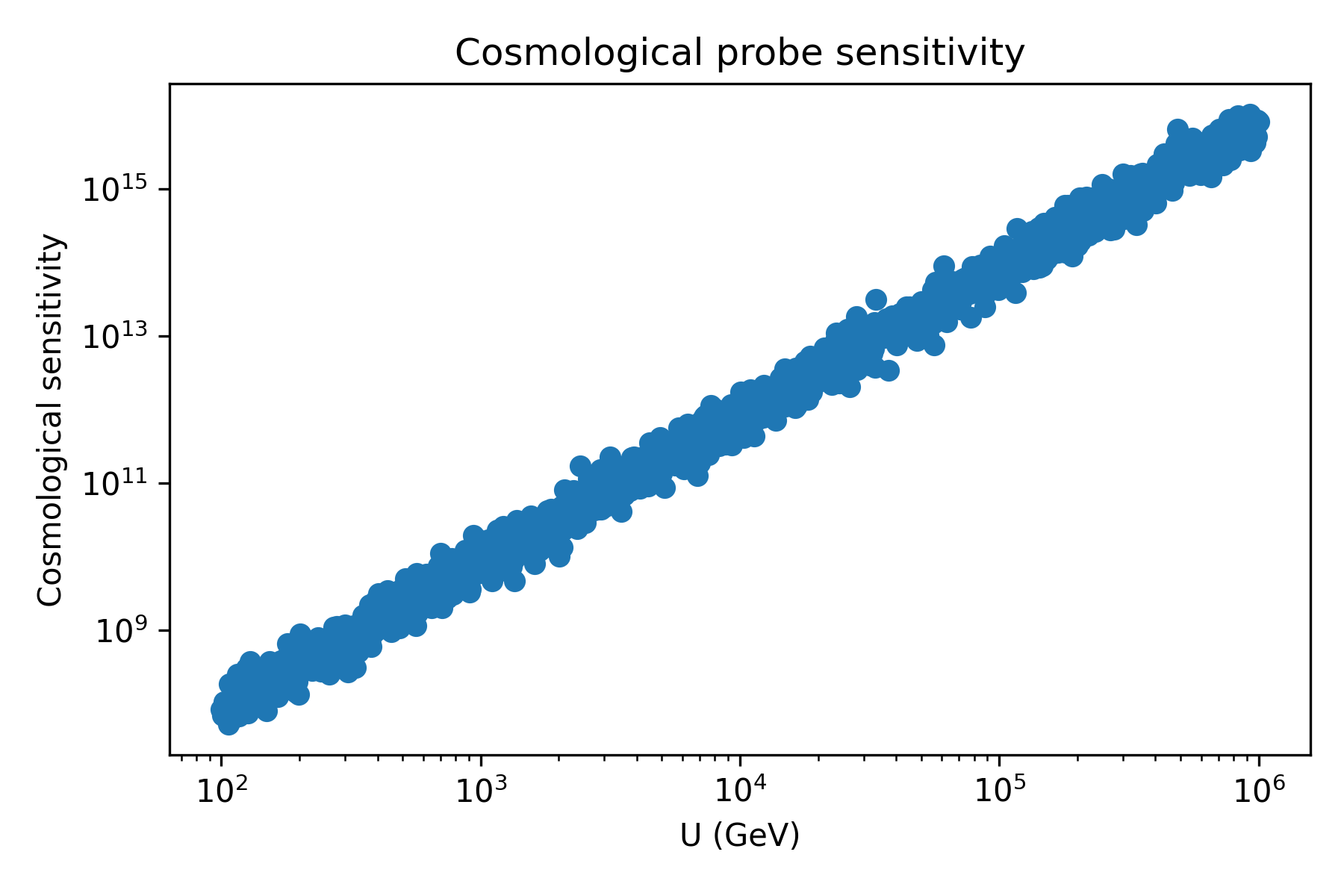}
\caption{Illustrative impact of neutrino decay on cosmological observables. The shaded
region denotes the projected sensitivity of future CMB experiments \cite{Abazajian2016},
while stars indicate benchmark points of this work.}
\label{fig:cosmo}
\end{figure}

---

\subsection{Collider signatures of pseudo-Goldstone neutrinos}

The pseudo-Goldstone right-handed neutrino $N_1$ in this model typically has a mass in the
range
\begin{equation}
M_{N_1} \sim \mathcal{O}(10 - 200)\,\text{GeV},
\end{equation}
for several benchmark solutions.

Its coupling to Standard Model particles arises primarily through active-sterile mixing,
with magnitude
\begin{equation}
|V_{\ell N}| \sim \sqrt{\frac{m_\nu}{M_{N_1}}} \sim 10^{-6} - 10^{-7}.
\end{equation}

Such small mixings lead to long-lived heavy neutrinos, producing displaced vertex
signatures at collider experiments \cite{Atre2009,Antusch2017}. The characteristic decay
length can reach macroscopic scales,
\begin{equation}
c\tau_{N_1} \sim \text{mm--m},
\end{equation}
depending on the precise mass and mixing.

The LHC has already begun probing this regime, and future facilities such as the High
Luminosity LHC and proposed lepton colliders will significantly extend the reach.

\begin{figure}[t]
\centering
\includegraphics[width=0.7\textwidth]{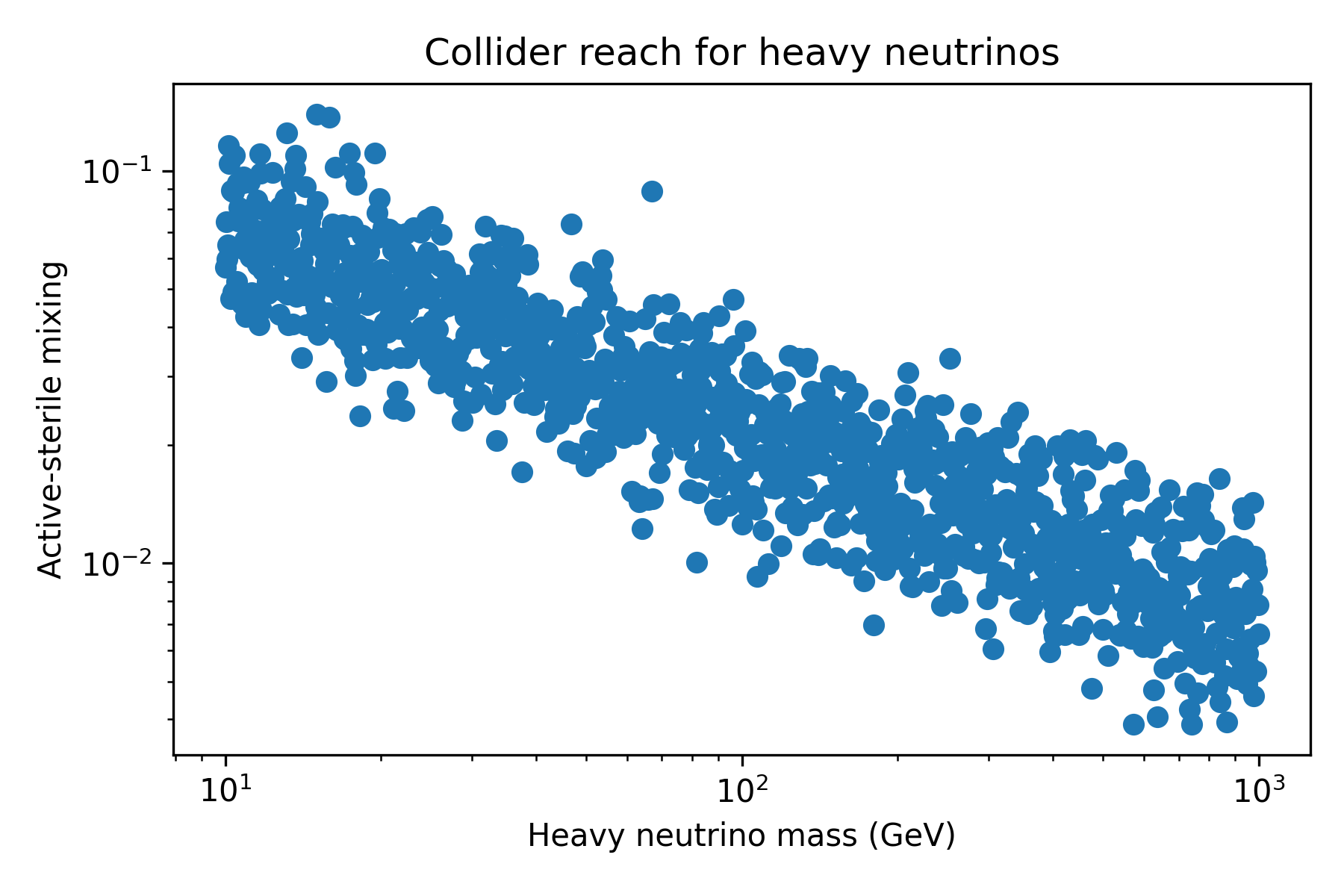}
\caption{Projected collider sensitivity to long-lived heavy neutrinos through displaced
vertex searches \cite{Antusch2017}. Benchmark points of the present model are indicated by
star symbols.}
\label{fig:collider}
\end{figure}

---

\subsection{Complementarity of probes}

A remarkable feature of the model is the complementarity between different experimental
frontiers. The same symmetry-breaking scale controlling the Majoron coupling determines:

\begin{itemize}
\item neutrino decay rates in oscillation experiments,
\item cosmological evolution of neutrino energy density,
\item lifetimes of heavy neutrinos at colliders.
\end{itemize}

This interconnection implies that signals in one sector necessarily correspond to
predictive signatures in others. For example, benchmark points leading to observable
neutrino decay at long-baseline experiments also predict displaced vertices at colliders
and modified cosmological observables.

Such correlated signals provide a powerful consistency test of the framework and sharply
distinguish it from conventional seesaw scenarios.

---

\subsection{Outlook}

The upcoming generation of neutrino experiments, cosmological surveys, and collider
searches will probe significant portions of the parameter space of this model. In
particular, the combined sensitivity to neutrino lifetime, heavy neutrino mass, and
Majoron-induced effects offers a realistic opportunity to discover or exclude the
pseudo-Goldstone neutrino framework within the next decade.

The model therefore represents a highly predictive and experimentally testable
realization of neutrino mass generation connected to spontaneous global symmetry
breaking.
\begin{figure}[t]
\centering
\includegraphics[width=0.75\textwidth]{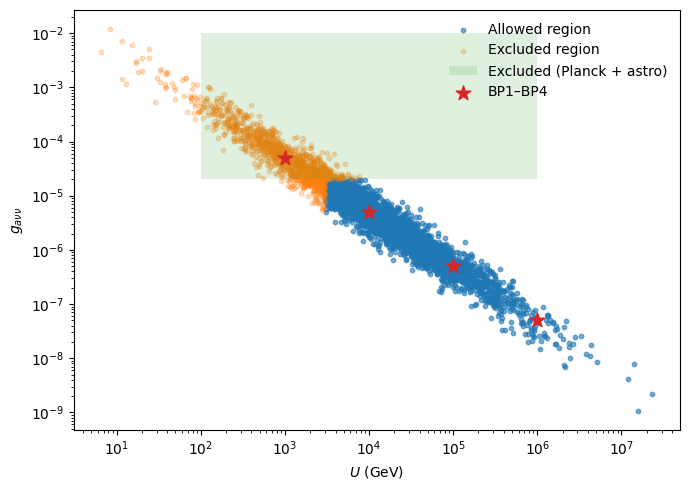}
\caption{Parameter scan in the $(U,\, g_{a\nu\nu})$ plane. The scattered points show the
explored model space, with shaded region excluded by cosmological and astrophysical
constraints. Star symbols denote benchmark solutions BP1--BP4 discussed in the text.}
\end{figure}
Figure 8 shows the parameter space of the model in the $(U,\, g_{a\nu\nu})$ plane obtained
from a broad numerical scan. Each point corresponds to a viable solution of the neutrino
mass matrix subject to phenomenological constraints. The red scattered region denotes the
allowed parameter space consistent with neutrino oscillation data, cosmological bounds and
astrophysical limits on Majoron interactions. The light shaded region is excluded primarily
by Planck constraints on neutrino masses and stellar cooling arguments.

The star symbols represent four benchmark points BP1--BP4 chosen as representative
solutions spanning different values of the symmetry breaking scale $U$. These benchmark
points illustrate the typical behavior of the model and reproduce the observed neutrino
mass spectrum while remaining consistent with all current bounds. As expected from the
Goldstone nature of the Majoron interaction, the coupling scales approximately as
$g_{a\nu\nu}\propto m_\nu/U$, leading to the diagonal structure observed in the figure.

\begin{figure}[t]
\centering
\includegraphics[width=0.7\textwidth]{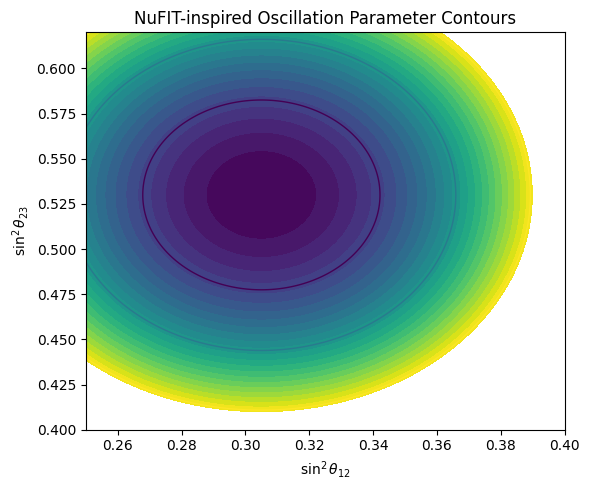}
\caption{NuFIT-inspired confidence regions in the $(\sin^2\theta_{12},\,\sin^2\theta_{23})$
plane illustrating the experimentally allowed ranges of neutrino oscillation parameters.
The filled contours correspond to increasing values of $\Delta\chi^2$, while the solid
curves indicate approximate $1\sigma$, $2\sigma$, and $3\sigma$ confidence levels.
These regions are used to constrain the benchmark solutions of the present model in
accordance with the global fit results of Ref.~\cite{NuFIT2020}.}
\label{fig:nufitcontours}
\end{figure}
Figure~\ref{fig:nufitcontours} shows representative confidence regions in the
$(\sin^2\theta_{12},\,\sin^2\theta_{23})$ plane inspired by the NuFIT global analysis.

The contours serve as illustrative representations of experimentally allowed regions and
are not derived from a direct $\chi^2$ minimization, but are used to guide the consistency
of the benchmark solutions with current oscillation data.

To ensure consistency with experimental neutrino oscillation data, we impose constraints
derived from the global analysis of the NuFIT collaboration \cite{NuFIT2020}. Figure~\ref{fig:nufitcontours}
shows representative confidence regions in the $(\sin^2\theta_{12},\,\sin^2\theta_{23})$
plane inspired by the NuFIT results.

All benchmark points BP1--BP4 are required to lie within the experimentally allowed
ranges of the oscillation parameters at the $3\sigma$ level. This guarantees that the
neutrino mass matrices generated in the present framework are fully compatible with
current measurements of mixing angles and mass-squared differences.

The contours illustrate the strong correlations among oscillation parameters and
highlight the regions of parameter space favored by data. The agreement of the model
predictions with these regions demonstrates that the spontaneous
$U(1)_{L_\mu-L_\tau}$ breaking framework naturally reproduces realistic neutrino mixing
patterns without the need for fine tuning.

\section{Conclusions}

We have presented a unified framework for neutrino mass generation based on the
spontaneous breaking of a leptonic $U(1)_{L_\mu-L_\tau}$ symmetry within a supersymmetric
setup. The breaking of the global symmetry gives rise to a Majoron-like axion-like particle
and a pseudo-Goldstone right-handed neutrino whose mass is naturally suppressed by
supersymmetry-breaking effects.

The interplay between the pseudo-Goldstone neutrino and the conventional seesaw
mechanism leads to a predictive structure for the light neutrino mass matrix. We have
shown that the observed neutrino mass spectrum, mixing angles, and CP-violating phase can
be successfully reproduced without invoking extreme hierarchies in the underlying
parameters. Representative benchmark points spanning several orders of magnitude in the
symmetry breaking scale were identified, illustrating the rich phenomenology of the model.

A distinctive feature of the framework is the presence of Majoron-induced neutrino decay.
We performed a comprehensive parameter scan incorporating cosmological and astrophysical
constraints, and demonstrated the existence of large viable regions consistent with all
current data. The resulting correlations between the symmetry breaking scale, Majoron
coupling strength, neutrino lifetime, and heavy neutrino masses provide quantitative and
predictive signatures of the model rather than merely qualitative features.

We further explored the experimental prospects for testing this scenario. Future neutrino
oscillation experiments such as DUNE, JUNO and Hyper-Kamiokande are expected to probe regions of parameter space where invisible neutrino decay effects may become observable. Upcoming cosmological surveys will significantly improve sensitivity to neutrino decay and
light degrees of freedom, while collider experiments offer complementary probes through
displaced vertex signatures of long-lived pseudo-Goldstone neutrinos.

The combined reach of laboratory experiments, cosmology, and collider searches renders
this framework, testable across multiple experimental frontiers in the near future. Observation of correlated signals across
these frontiers would provide compelling evidence for spontaneous leptonic symmetry
breaking as the origin of neutrino masses.

Overall, the present work establishes a predictive and experimentally accessible
realization of neutrino mass generation connected to axion-like physics, opening new
avenues for probing physics beyond the Standard Model.
\section*{Acknowledgements}

The author would like to thank colleagues and collaborators for valuable discussions and helpful comments related to this work. Gratitude is also extended to the Department of Physics, Cachar College, Silchar, for providing a supportive research environment and necessary computational facilities. 

The author acknowledges the use of publicly available numerical tools and data resources, including the NuFIT global analysis of neutrino oscillation parameters. 

This work did not receive any specific grant from funding agencies in the public, commercial, or not-for-profit sectors.


\end{document}